\RequirePackage{fix-cm}
\documentclass[twocolumn]{svjour3}          
\smartqed  
\usepackage{graphicx}
\begin{document}

\title{Vortex Structure and Anisotropic Superconducting Gaps in Ba[Fe(Ni)]$_2$As$_2$}

\titlerunning{Vortex Structure and Anisotropic Superconducting Gaps in BFNA}

\author{T.E. Kuzmicheva \and
V.A. Vlasenko \and
S.Yu. Gavrilkin \and
S.A. Kuzmichev \and
K.S. Pervakov \and
I.V. Roshchina \and
V.M. Pudalov
}

\institute{T.E. Kuzmicheva \and
S.Yu. Gavrilkin \and
K.S. Pervakov \and
V.A. Vlasenko \and
V.M. Pudalov \at
              P.N. Lebedev Physical Institute of the RAS, 119991 Moscow, Russia \\
              Tel.: +7-499-132-60-48\\
              \email{kute@sci.lebedev.ru}           
           \and
           S.A. Kuzmichev \and
           I.V. Roshchina \at
              M.V. Lomonosov Moscow State University, 119991 Moscow, Russia \\
}

\date{Received: date / Accepted: date}

\maketitle

\begin{abstract}

We studied nearly optimally Ni-substituted BaFe$_{2-x}$Ni$_x$As$_2$ (BFNA) single crystals with $T_C \approx 18.5$\,K. In irreversible magnetization measurements, we determined the field dependence of the critical-current density and discuss the nature of observed strong bulk pinning. Using intrinsic multiple Andreev reflections effect (IMARE) spectroscopy, we directly determine two distinct superconducting gaps and resolve their moderate anisotropy in the momentum space. The BCS-ratio for the large gap $2\Delta_L/k_BT_C > 4.1$ evidences for a strong coupling in the $\Delta_L$-bands.

\end{abstract}

\section{Introduction}
The BaFe$_2$As$_2$ family (Ba-122) is currently the most intensively studied among the pnictide superconductors \cite{Kamihara} due to rather high critical temperatures up to $T_C \approx 38$\,K \cite{Rotter} and quite simple growth of large single crystals. Ba-122 possess a layered crystal structure with Fe--As blocks separated with Ba-based spacers along the crystallographic $c$-direction. Because of high critical fields and critical current density, the Ba-122 superconductors are very perspective for high-field applications. Significant efforts have been devoted to understanding their properties. Studying of vortex matter in Ba-122 compounds may provide an opportunity to understand the crossover between low temperature and high temperature superconductors. Among the 122 pnictide superconductors one of the most studied compounds is Co-doped BaFe$_2$As$_2$ \cite{Sefat} where high flux creep rates and a transition from collective to plastic creep have been reported. The pinning behaviour in Co-doped BaFe$_2$As$_2$ is rather complex, and different sources of pinning have been discussed: grain boundaries \cite{Kalisky} and nanoscale variations of $T_C$ and/or the inhomogeneous distribution of dopant atoms \cite{Demirdis,Haberkorn}. In this paper, we study superconducting properties of the nearly optimum BaFe$_{2-x}$Ni$_x$As$_2$ single crystals. We found that strong bulk pinning in BFNA is similar to that in Co-doped Ba-122 and arises due to random point-like nanoparticles. In the high-field range, pinning is caused by combination of several mechanisms related to the point defects.

Several bands crossing the Fermi level produce quasi-two-dimensional Fermi surface sheets, the electron-like near the M point, and the hole-like near the $\Gamma$ point of the Brillouin zone \cite{Evtushinsky}. Despite the intensive studies of the 122-family, the available experimental data are contradictive \cite{Seidel}. The very likely reason for this is a complex and unconventional nature of superconductivity, in particular, a theoretically supposed \cite{Maiti,Kontani} anisotropy of superconducting properties in momentum space. In general, considering the importance of various intra- and interorbital interactions, theoretical studies offer two basic pairing models, the so called $s^{++}$ and $s^{\pm}$. In the $s^{\pm}$ model, superconductivity arises through a strong interband coupling via spin fluctuations \cite{Maiti}. In contrast, recent $s^{++}$ calculations take into account two fundamental pairings, via spin and orbital fluctuations \cite{Kontani}. The competition of these mechanisms could result in anisotropic or even nodal order parameter \cite{Kontani}. Consequently, the gap symmetry and the pinning symmetry might be capable to reveal the underlying pairing mechanism.

Here we present detailed studies of superconducting properties of nearly optimal Ni-substituted BaFe$_{2-x}$Ni$_x$As$_2$ single crystals with $T_C \approx 18.5$\,K. Using irreversible magnetization data, we discuss pinning effects in Ba-122. In Andreev spectroscopy studies, we directly determine the structure of superconducting order parameter.

\section{Experimental details}

BaFe$_{2-x}$Ni$_x$As$_2$ single crystals 
 with x=0.093 and x=0.1 and critical temperature $T_C \approx 18.5$\,K were grown using the self-flux method. High purity Ba, FeAs and NiAs were mixed in $1: 5(1-x):5x$ molar ratio, placed in alumina crucible and sealed in a quartz tube under 0.2\,bar argon pressure. The ampoule was heated up to $1200^o$\,C and then cooled down to $1000^o$\,C at 2$^o$\,C/h rate. The grown crystals are up to $4 \times 2 \times 0.2$\,mm$^3$ in size and possess a single superconducting phase.
The irreversible magnetization $M(H,T)$ measurements were performed using the PPMS vibrating sample magnetometer in fields up to 9\,T applied along the crystal $c$-axis. The typical field sweep rate was 100\,Oe/s.

The superconducting order parameter was probed using multiple Andreev reflections effect (MARE) spectroscopy \cite{Pon_IMARE,SSC2004,JETPL2004,EPL}. In ballistic superconductor-normal metal-superconductor (SnS) junction (the junction dimension $2a$ is less than the carrier mean free path $l$), MARE manifests itself as a pronounced excess conductance at low bias voltages (so called ``foot''), and a subharmonic gap structure (SGS). In case of high-transparency SnS-junction, the SGS is a series of dynamic conductance dips at positions $V_n = 2\Delta/en$, where $\Delta$ is the superconducting gap, $e$ \textemdash elementary charge, and $n = 1, 2, \dots$ \textemdash subharmonic order \cite{OTBK,Averin,Arnold,Kummel}. Therefore, the gap may be directly determined from the positions of Andreev subharmonics at $0 < T < T_C$ \cite{Kummel}. Obviously, two distinct gaps would cause two SGS in a dynamic conductance spectrum. An anisotropy of the superconducting gap strongly affects the shape of Andreev subharmonics \cite{Devereaux,Li2013,BJ}. Isotropic ($s$-wave) gap produces high-intensive and symmetrical dips, while $d$-wave or fully anisotropic $s$-wave gap make subharmonics poorly visible and strongly asymmetric. Anisotropic gap with $\Delta(\theta) \sim cos(4\theta)$ angle distribution in the $k-$space (which is very likely the case for Ba-122 \cite{Maiti}) for $c$-direction transport causes doublet-like features with two minima connected by an arch, which positions correspond to the higher and lower extremes of the gap angular distribution \cite{BJ}.

High-quality SnS-contacts were formed by a ``break-junction'' technique \cite{BJ,Moreland}. The single crystal plate was attached to a springy sample holder using pads of In-Ga paste, and then cooled down to $T = 4.2$\,K. In the cryogenic environment, a gentle mechanical curving of the sample holder produced a cryogenic cleavage of the sample, thus creating two superconducting banks separated with a weak link (ScS-contact, $c$ \textemdash constriction). In Ba-122, the constriction usually acts as a thin normal metal, making it possible to observe MARE in a ballistic SnS-contact \cite{OTBK,Arnold,Kummel,Andreev}. Since the superconducting banks are located at a tiny distance during the experiment, our technique preserve the crack from impurity penetration and provides clean cryogenic surfaces to probe the gap(s) magnitude almost unaffected by surface defects. When varying precisely the curvature of the sample holder, two cryogenic surfaces slide apart along the $ab$-plane, forming up to several tens of ScS-contacts with various dimension and resistance. This helps to collect a large amount of data with one and the same sample in order to check data reproducibility to be aware of dimensional effects.

Another unique feature of the ``break-junction'' is the formation of ScSc-\dots-S arrays \cite{Pon_IMARE,EPL,FPS13} when probing a layered sample. The layered single crystal usually exfoliates along the $ab$-planes with the formation of steps and terraces along the $c$-direction, where an intrinsic multiple Andreev reflections effect (IMARE) occurs. Strictly speaking, IMARE is similar to intrinsic Josephson effect \cite{Nakamura}, firstly observed in cuprates \cite{Pon_IJE}. When considering an Andreev array as a sequence of $m$ identical SnS-junctions, the position of SGS scales with $m$: $V_n = \frac{2\Delta}{en} \times m, ~~~~~ n,m = 1,2\dots$. In order to determine $m$ and the gap(s) value, one should find such natural numbers which scale the I(V) and dI(V)/dV curves for various arrays onto each other, or to to achieve the same position of gap features with those for a single-junction spectrum. In such array, the contribution of bulk effects well exceeds that of the surface influence \cite{EPL,BJ}. Strictly speaking, the IMARE spectroscopy is currently the only technique probing the $bulk$ values of superconducting gap(s) $locally$ (within the contact area of $10 \textendash 50$\,nm) \cite{BJ}.

Thr dI(V)/dV curves were directly measured using a current source and a standard modulation technique \cite{Pon_LOFA}. Obviously, in our case the current flows along the $c$-direction, and the corresponding velocity components of the charge carriers are lying almost in-plane, $v_c \ll v_{a,b}$ (due to the quasi-two-dimensionality of the Ba-122 compounds). This property enables to probe the gap anisotropy in the $ab$-plane \cite{BJ}. In case of several conducting bands, for our contact geometry, several conduction channels contribute in parallel and the respective gaps may be quantified independently.

\section{Irreversible magnetization}

Figure 1 shows typical isothermal magnetization M(H) loops (MHL) measured at several temperatures from 4 to 18\,K in magnetic fields $H \parallel c$ up to 9\,T for BaFe$_{1.907}$Ni$_{0.093}$As$_2$ single crystal. The almost perfect symmetry of the MHLs indicates domination  of the bulk pinning and exhibits no magnetic background. Therefore, we conclude our sample contains negligible amount of magnetic impurities \cite{Ba} and shows bulk vortex pinning. Additionally, in MHLs of BFNA single crystal one can see the second magnetization peak (SMP) \cite{Refb} or the ``fish-tail''. The observed fish-tail for the present Ni-doped single crystal is similar to that seen earlier for Co, Na, and P-doped Ba-122 single crystals \cite{Refc,Refd}. This implies the SMP in the MHL is common to the Ba-122 superconductors \cite{Refe}.

\begin{figure}
\includegraphics [width=0.5\textwidth]{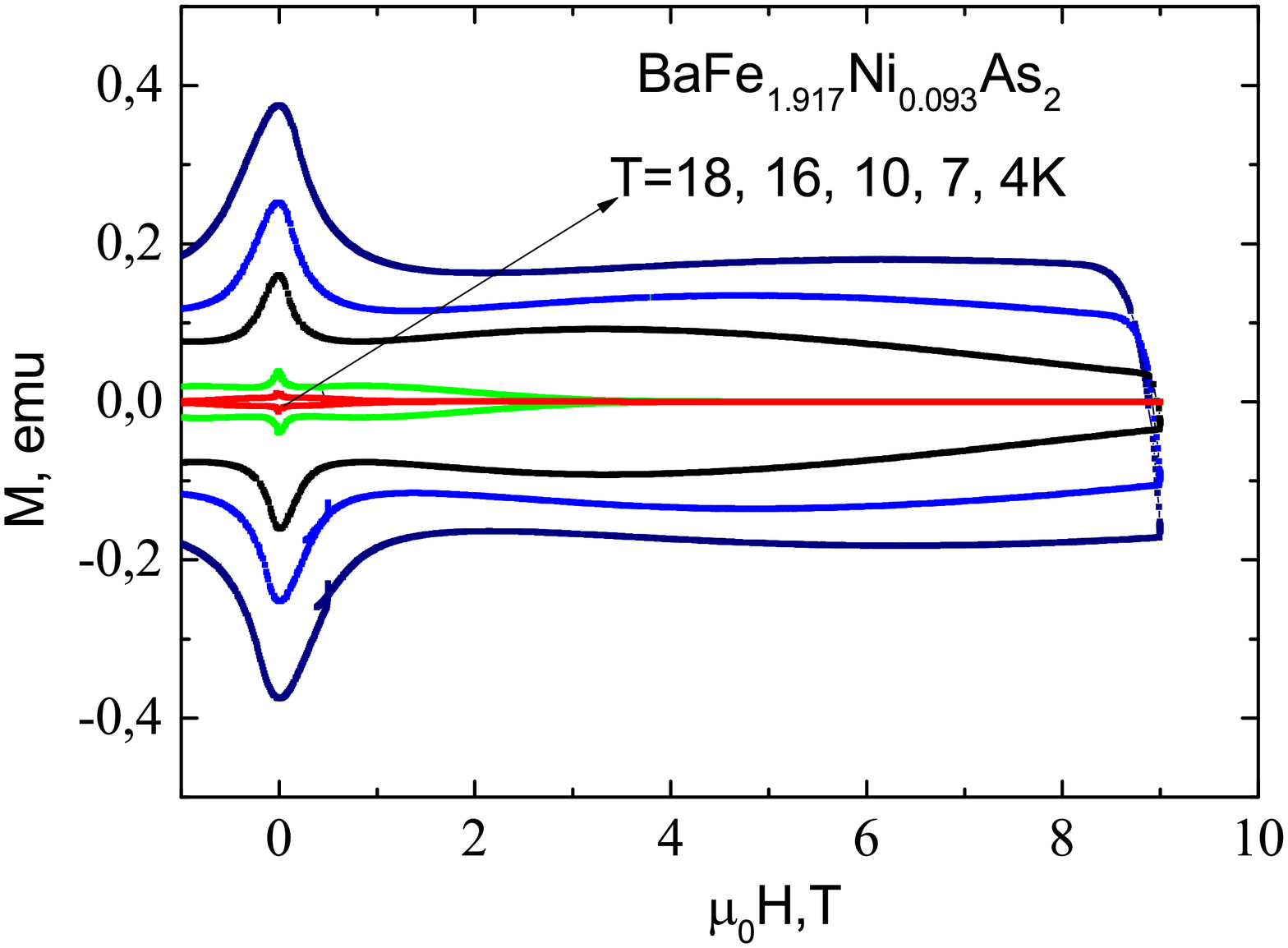}
\caption{Magnetization hysteresis loops (MHL) of BFNA single crystal at various temperatures.}
\label{fig:1}
\end{figure}

According to the Bean model for current distribution \cite{Reff}, $J_c$ is given by the formula: $J_c = 20M/{a(1 - a/3b)}$, where $M = M_{down} - M_{up}$, $a$ and $b$ are the sample transverse dimensions with $a < b$. We calculate $J_c$ from MHLs using the Bean model with effective sample dimensions  $2 \times 1.5 \times 0.185$\,mm$^3$. Figure 2 shows on the log-log plot the field dependence of $J_c$ obtained from the MHLs data.

Haberkorn et al. \cite{Refg} has shown that J$_c(H)$ behaviour exhibits several regimes: (I) low-field part associated with the single vortex regime; (II) a power-law dependence $J_c$ $\sim H
^{−a}$, associated with strong pinning centers; (III) a regime related with the fishtail; and (IV) a fast drop in $J_c(H)$ where the vortex dynamics changes to plastic. As one can clearly see several marked area in Fig.2 are related with different pinning regimes. Regime I is observed at low fields up to 150\,Oe. The power law behaviour $J_c \sim H^{p}$ extends up to 2\,T. The obtained values of $p$ were from 0.45 to 0.58 for temperature interval between 18 to 4\,K respectively. The value of the exponent is in good agreement with the theoretical prediction of $\sim H^{5/8}$. Such value of the exponent $p$ indicates strong vortex pinning \cite{Refl}. The III regime behaviour is related with crossover or/and correlation between strong pinning by defects and weak intrinsic pinning by large amount of centers \textemdash the so-called ``caging'' effect (CE). This leads to the conclusion that pinning in fields $< 2$\,T originates from different mechanisms. The IV region is associated with disruption of magnetic vortices from pinning centers, which leads to the dissipation of energy and a rapid drop of the critical current.
The onset of the IV regime is the limit of practical applicability of the superconductor. It should be noted that almost the same behavior of $J_c$ is observed in optimally doped BFNA.

\begin{figure}
\includegraphics[width=0.5\textwidth]{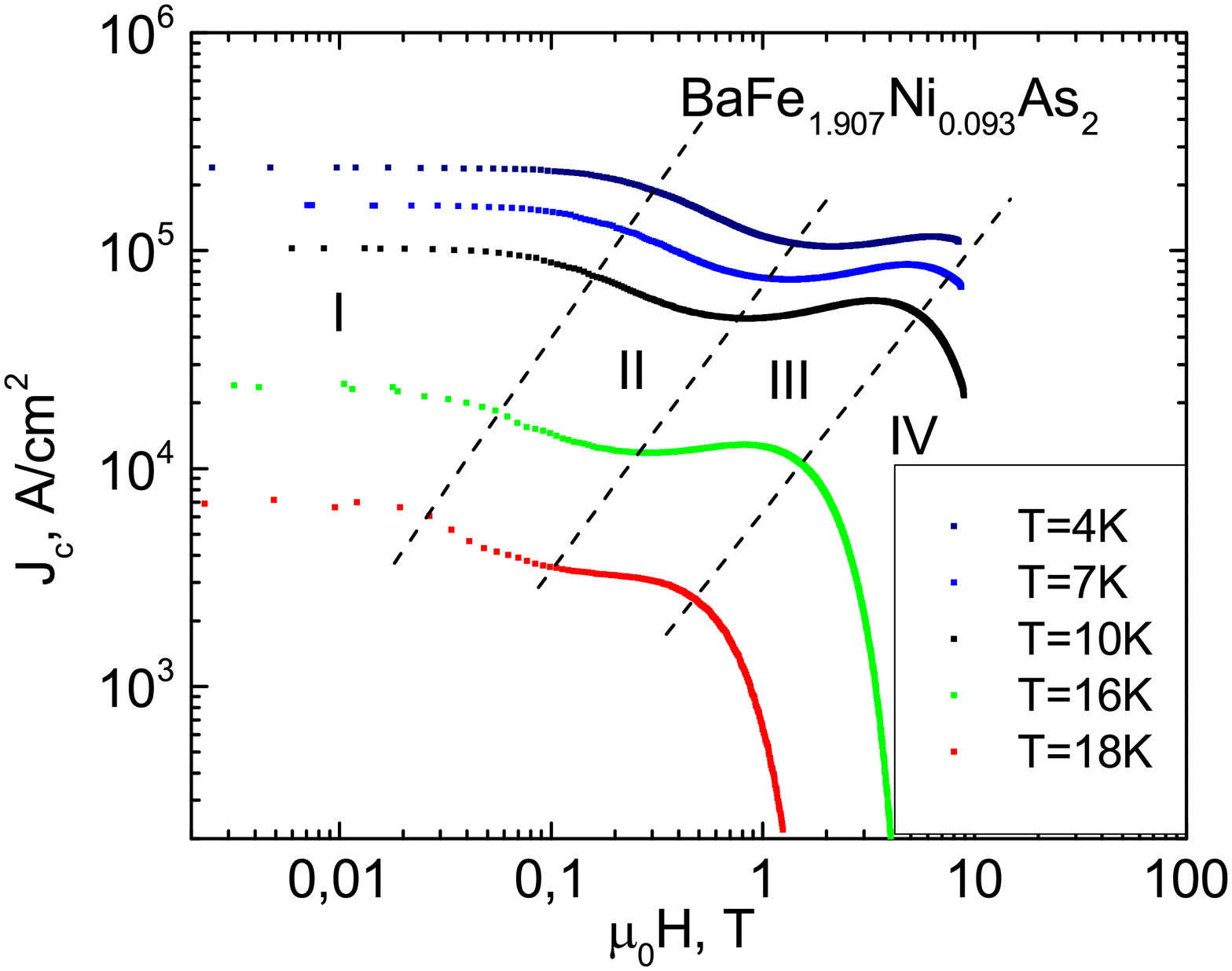}
\caption{Critical-current density ($J_c$) as a function of magnetic field (H) for BFNA single crystal.}
\label{fig:2}
\end{figure}

\section{Andreev spectroscopy}

Figure 3 shows current-voltage characteristic (CVC) and its derivative for a ScS-array formed in BFNA single crystal. The bias voltage axis was scaled with the natural $m$ to a single-junction spectrum, therefore, all the data corresponds to a single junction. The pronounced foot in CVC is typical for a clean SnS-mode \cite{OTBK,Averin,Arnold,Kummel}. To check whether the contact is ballistic, we take the normal-state bulk resistivity $\rho(22 {\rm K}) \approx 1.9 \times 10^4$\,${\rm \Omega} \cdot$cm for the sample under study, and the contact resistance $R \approx 200\,{\rm \Omega}$ (see Fig. 3) and plug them into the Sharvin formula $a = \sqrt{\frac{4}{3\pi}\frac{\rho l}{a^2}}$ \cite{Sharvin}. Given the product of the bulk resistivity and the carrier mean free path for Ba-122 determined elsewhere \cite{Xu} $\rho^{ab} l \approx 1.7 \times 10^{-9}\,{\rm \Omega} \cdot $cm, we immediately get $2a \approx 36\,{\rm nm} < l \approx 90\,{\rm nm}$, thus proving the contact is ballistic and favourable for MARE observation. Note the both parts of CVC (the positive and negative current) in Fig. 3 are symmetrical and non-hysteretic. In the dI(V)/dV spectrum (red line), two Andreev subharmonics for the large gap (observable in bias voltages $V_1 = 2\Delta_L/e \approx 9.2$\,mV, and $V_2 = \Delta_L/e \approx 4.6$\,mV) are well pronounced. The position of these dips determine the large gap $\Delta_L \approx 4.6$\,meV. Taking into account a strongly asymmetrical shape of both dips, one may suppose a gap anisotropy in the $k-$space \cite{Devereaux,BJ}. If it is the case, the aforementioned dips at $V_1$, $V_2$ correspond to larger extremum of the gap distribution $\Delta_L^{max} \approx 4.6$\,meV. The features located at 6.6 and 3.3\,mV are very likely the subharmonics for the lower extremum $\Delta_L^{min} \approx 3.3$\,meV. Given this assumption, the large gap anisotropy may be estimated as $A = [1- \Delta_L^{min} / \Delta_L^{max}] \cdot 100\% \approx 30 \%$.
The complex fine structure of the observed Andreev dips is caused by the $ab$-plane gap distribution; it seems to differ from the simple $\Delta(\theta) \sim cos(4\theta)$ predicted e.g. in \cite{Kontani}. In the inset of Fig. 3, we show the position of $\Delta_L$ doublets $V_n$ versus their inverse number $1/n$. The linear dependence crossing the origin proves these features are related to one and the same SGS. The above figures demonstrate the observation of IMARE in BFNA single crystals.

\begin{figure}
\includegraphics[width=0.5\textwidth]{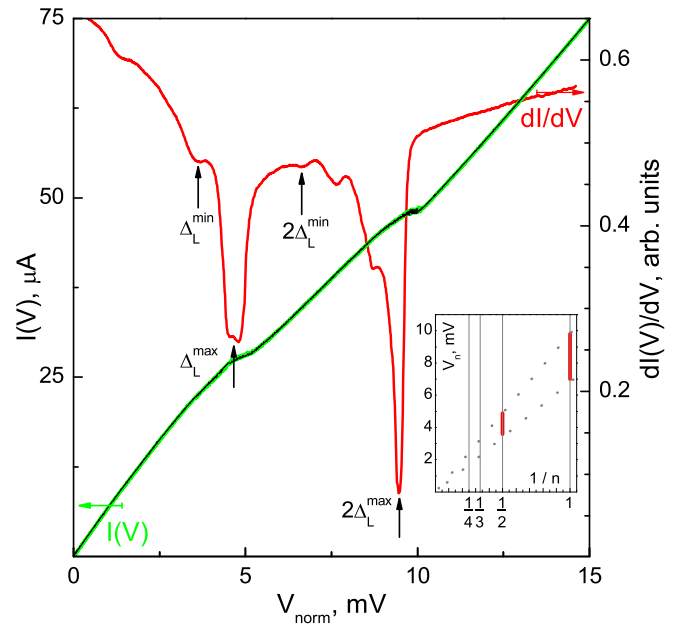}
\caption{Non-hysteretic current-voltage characteristic (green and black lines for positive-bias part and reversed negative-bias parts, correspondingly) and dynamic conductance spectrum for Andreev array in BFNA single crystal. Andreev subharmonics of the large gap are shown by arrows. The inset shows the linear dependence of subharmonics position versus $1/n$ for $\Delta_L$, the length of the red dashes indicate a possible gap anisotropy $\sim30\%$.
}
\end{figure}

Two Andreev spectra of the SnS-arrays formed sequentially (after a mechanical readjustment) in one and the same BFNA sample are shown in Fig. 4. Despite the different contact resistance, the spectra look similarly and possess features for both the large and the small gap. The main doublet-like dip located at $V_1 \approx 6.4 \textendash 8.5$\,mV is nearly twice wider than the second feature for the large gap at $V_2 \approx 4.4 \textendash 3.2$\,mV. Together, these dips are the $n = 1, 2$ subharmonics for the anisotropic large gap $\Delta_L \approx 3.2 \textendash 4.4$\,meV, whereas their width corresponds to $\sim 30 \%$ anisotropy of the large gap.
The latter value is close to that determined using data of Fig. 3. The intensive doublets at $2.4 \textendash 3.2$\,mV do not correspond to the large gap SGS, and therefore, may be interpreted as related to the small gap with moderate anisotropy $\Delta_S \approx 1.2 \textendash 1.6$\,meV ($\sim 25\%$ anisotropy).
Note that for the two spectra shown in Fig. 4, the position of both-gap features coincide. The gap values remain still unchanged during the contact readjustment confirming both a high quality of cryogenic interfaces and a high homogeneity of the single crystal under study. The uniform distribution of dopants within the typical contact area reduces a contribution of $\delta T_C$-pinning mechanism (associated with a spatial variation of the superconducting transition temperature $T_C$ throughout the sample). Therefore, pinning in the studied Ba-122 samples may be caused by randomly distributed point-like pinning centers ($\delta l$-pinning) at a sub-nanometer scale \cite{Blatter}.

\begin{figure}
\includegraphics[width=0.5\textwidth]{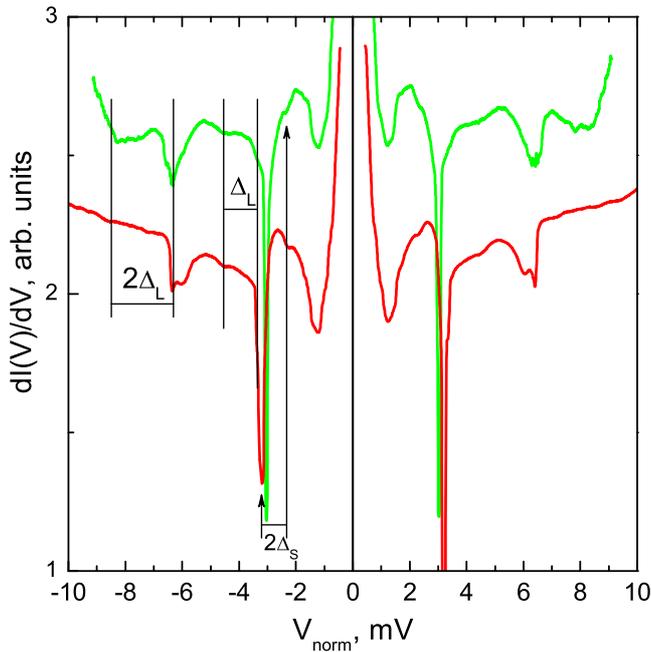}
\caption{Normalized dynamic conductance spectra of Andreev arrays (with different number of junctions) formed in BFNA single crystal. Doublet-like subharmonics for the anisotropic large gap $\Delta_L = 3.2 \textendash 4.4$\,meV ($\sim 30\%$ anisotropy) are shown by thin black lines. Arrows indicate the main doublet features of the small gap $\Delta_S = 1.2 \textendash 1.6$\,meV ($\sim 25\%$ anisotropy).}
\end{figure}

From the IMARE spectra, we determined the BCS-ratio $2\Delta_L/k_BT_C \approx 4.1 \textendash 5.8$. It is similar to that determined earlier in IMARE experiments with K-doped Ba-122 single crystals \cite{Ba,BKFA} and exceeds the standard weak-coupling limit 3.52, thus proving a strong coupling. Taking into account the moderate gap anisotropy determined here, the BCS-ratio well covers all the range of $2\Delta_L/k_BT_C$ estimated in specific heat and $H_{C1}$ measurements with the BFNA samples from the same batch \cite{Ba2016}, and in infrared spectroscopy and magnetization experiments with nearly optimal K-doped Ba-122 single crystals with $T_C = 35 \textendash 37$\,K \cite{Ba,BKFA}. It is also consistent with other published data for Ba-122 \cite{Evtushinsky,Hardy,Pramanik,Khasanov,Yin}. The BCS-ratio for the small gap is $2\Delta_S/k_BT_C \approx 1.5 \textendash 2.0$ and lies well below the weak-coupling limit, thus proving the presence of the $k$-space proximity effect and substantial interband interaction.

\section{Conclusions}
We have studied the field dependence of the critical-current density and the structure of superconducting order parameter in nearly optimally doped BaFe$_{2-x}$Ni$_x$As$_2$ ($x = 0.07, 0.1$) single crystals. Our results indicate the strong bulk pinning in BFNA by random nanoparticles dominates at low fields. At higher fields, the strong pinning may originate from a combination of several mechanisms related to point defects. Using intrinsic multiple Andreev reflections effect (IMARE) observed in BFNA, we directly determine the values of two distinct superconducting gaps with a moderate anisotropy in a $k$-space: $\Delta_L \approx 3.3 \textendash 4.5$\,meV (the range corresponds to $\sim 30\%$ anisotropy), $\Delta_S \approx 1.2 \textendash 1.6$\,meV ($\sim 25\%$ anisotropy).
The BCS-ratio for the large gap $2\Delta_L/k_BT_C \approx 4.1 \textendash 5.8 > 3.52$ is close to that for K-doped Ba-122 \cite{Ba} and evidences for a strong coupling in the $\Delta_L$-bands.

\begin{acknowledgements}
We thank M. Abdel-Hafiez, Y. Chen, H. Kontani, Yu.F. Eltsev for useful discussions and the samples presented. The work was partially supported by the Council of the President of the Russian Federation for Support of Young Scientists (project no. MK-5699.2016.2). VMP acknowledges support from RSCF (no. 16-42-01100). The measurements were partly done using research equipment of the Shared Facilities Center at LPI.
\end{acknowledgements}

\end{document}